\documentclass[twocolumn,showpacs,superscriptaddress,amsmath,amssymb,pre]{revtex4-1}
\usepackage{dcolumn}
\usepackage{bm}
\usepackage[dvips]{graphicx}
\usepackage{wrapfig,subfigure}
\usepackage{color}

\begin{document}

\title{Singular probability distribution of shot-noise driven systems}

\author{Akihisa Ichiki}
\author{Yukihiro Tadokoro}
\affiliation{Toyota Central R\&D Labs., Inc., Nagakute, Aichi 480-1192, Japan}
\author{M. I. Dykman}
\affiliation{Department of Physics and Astronomy, Michigan State University, East Lansing, Michigan 48824, USA}

\date{\today}

\begin{abstract}
We study the stationary probability distribution of a system driven by shot noise. We find that both in the overdamped and underdamped regime, the coordinate distribution displays power-law singularities in its central part. For sufficiently low rate of noise pulses they correspond to distribution peaks. We find the positions of the peaks and the corresponding exponents. In the underdamped regime the peak positions are given by a geometric progression. The energy distribution in this case also displays multiple peaks with positions given by a geometric progression. Such structure is a signature of the shot-noise induced fluctuations. The analytical results are in excellent agreement with numerical simulations.
\end{abstract}


\maketitle

\section{Introduction}

Shot noise is an important source of fluctuations in dynamical systems. The discreteness of the modulating force which underlies such noise can be due to the quantization of the electromagnetic radiation that drives the system or the discreteness of the electron charge or spin in the electric or spin current in the system. The features of the system dynamics provide means for identifying the noise statistics, which is sensitive to the microscopic nature of the noise source. An example of using the dynamics for detecting non-Gaussian noise statistics is the recent theoretical and experimental work on noise-induced switching between coexisting stable states in Josephson junctions and mechanical resonators \cite{Tobiska2004,Pekola2004,Jordan2005,*Sukhorukov2007,Ankerhold2007a,Timofeev2007,Billings2008,*Billings2010,Grabert2008,LeMasne2009,Zou2012}.

The strong effect of the noise discreteness is easy to see for a strongly damped noise-driven system localized in a potential well. If the noise pulses are all of the same sign, the noise pushes the system only in one direction, and the stationary probability distribution of the system is equal to zero on the one side of its stable state. In the case of shot noise, the distribution turns out to be singular near the stable state both for the case of pulses with random exponentially distributed amplitude \cite{VanDenBroeck1983,Sancho1987,Baule2009,Romanczuk2012} and pulses of constant amplitude \cite{Dykman2010a}. Depending on the pulse rate relative to the system relaxation rate, it either displays a power-law divergence or goes to zero as a power law of the distance to the stable state.

In this paper we study the probability distribution for a shot (Poisson) noise-driven system with arbitrary damping. Our results extend from the limit of overdamped dynamics to underdamped dynamics, where the relaxation rate is small compared to the typical vibration frequency of the system. Examples of underdamped systems that are of interest for studying Poisson-noise induced fluctuations include Josephson junctions \cite{Sukhorukov2009}, nano-magnetic oscillators \cite{Dunn2012}, and high-Q nanomechanical resonators coupled to electron tunneling, to mention but a few; the problem attracted much attention recently in the context of the studies of radiation-pressure shot noise with optomechanical systems \cite{Borkje2010,Purdy2012}. We assume that the Poisson noise pulses have constant amplitude, which is relevant for most of the above systems.

One would expect that, for an underdamped system, the aforementioned singularity of the probability distribution at the stable state should disappear. Indeed, in this case the noise makes the system perform random vibrations. Therefore, it contrast to an overdamped system, the probability distribution is nonzero on the both sides of the stable state. However, we find that the power-law singularity at the stable state persists. 

Moreover, we find that the probability distribution of Poisson-noise driven underdamped systems can display multiple power-law singularities, with positions forming a geometric progression. To gain an additional insight into this unusual structure we look at the probability distribution of the system energy. We show that this distribution also has singularities. 

It is somewhat surprising that more is known about the tail of the distribution of Poisson-noise driven systems \cite{Sancho1987,Sukhorukov2007,Billings2008,Grabert2008,Zou2012,Dykman2010a,Sukhorukov2009,VandenBroeck1984,Masoliver1987} than about its central part, which is singular. Unless the noise is very strong, this central part is formed by the motion of the system near its stable state, which is generally described by linear equations of motion. This allows us to obtain the results in a closed form and to study the critical exponents that characterize the singularities of the distribution.

In Sec.~\ref{sec:model} we present the model of a simple shot-noise driven system with inertia. The onset of multiple power-law singularities of the coordinate distribution in the overdamped and underdamped  regimes is studied in Secs.~\ref{sec:singularities} and \ref{sec:underdamped}, respectively. The onset of the singularities of the energy distribution of an underdamped system is considered in Sec.~\ref{sec:energy}. Section~\ref{sec:discussion} provides a qualitative insight into the onset of singularities and explains their positions. Section~\ref{sec:conclusions} contains concluding remarks.

\section{Model}
\label{sec:model}

We will consider a standard model where the dynamics of a system near its stable state is described by the Langevin equation
\begin{eqnarray}
\label{eq:Langevin}
&&\ddot q + 2\Gamma \dot q + \omega_0^2q=f_P(t), \\
&&f_P(t)=g\sum_n\delta (t-t_n).\nonumber
\end{eqnarray}
Here, $q$ is the system coordinate counted off from the equilibrium position, $\omega_0^2$ is the curvature of the effective confining potential, and $\Gamma$ is the viscous friction coefficient. The force $f_P(t)$ is a Poisson (shot) noise. It consists of short pulses which occur at random, the instants $t_n$ are uncorrelated. The average pulse rate is $\nu$. We assume that $f_P(t)$  is independent of the system coordinate, in agreement with our intention to keep the leading-order terms in $q$ and $\dot q$. 

The dynamics of the system (\ref{eq:Langevin}) is determined by two dimensionless parameters: the relative decay rate $\Gamma/\omega_0$ and the relative pulse rate $\nu/\Gamma$. The pulse area $g$ just scales the velocity $\dot q$. For concreteness, we assume $g>0$.

To obtain the stationary distribution of the system we assume that the initial state has decayed and write the coordinate as
\begin{eqnarray}
\label{eq:coordinate_explicit}
q(t)&=&\int_{-\infty}^t dt'f_P(t')\alpha(t-t'),\qquad  \alpha(t)= \frac{e^{\lambda_1t}-e^{\lambda_2t}} 
{\lambda_1 - \lambda_2},\nonumber\\
\lambda_{1,2}&=&-\Gamma\pm i(\omega_0^2-\Gamma^2)^{1/2},
\end{eqnarray}
In this equation  $\alpha(t)$ is the response function of the system.

Using the well-known expression for the characteristic functional of the Poisson noise \cite{FeynmanQM}, we can then write the stationary distribution of the system coordinate as
\begin{eqnarray}
\label{eq:distribution_general}
\rho(q)&=&\langle \delta\bigl(q-q(0)\bigr)\rangle = \int \frac{dk}{2\pi}\exp\left[-ikq-\frac{\nu}{\Gamma}\psi(k)\right],\nonumber\\
&&\psi(k)=\Gamma\int_0^{\infty}dt\left[1-e^{ikg\alpha(t)}\right].
\end{eqnarray}
From Eq.~(\ref{eq:distribution_general}), $\psi(-k)=\psi^*(k)$. Therefore, in much of the analysis of $\psi(k)$ we will focus on the region $k>0$.

\section{Power-law singularities of the coordinate distribution: Overdamped regime}
\label{sec:singularities}

Integration over the range of small $k$ in Eq.~(\ref{eq:distribution_general}) gives a smooth contribution to the distribution $\rho(q)$. Singular behavior of $\rho(q)$ is determined by the large-$k$ behavior of the function $\psi(k)$. We will discuss it in three limiting cases, which correspond to the overdamped and underdamped regimes, as well as in the critical regime where the dynamics changes from over- to underdamped.

In the overdamped regime the friction coefficient $\Gamma$ exceeds the frequency $\omega_0$. The system does not oscillate in the absence of noise. Both eigenvalues $\lambda_{1,2}$ in Eq.~(\ref{eq:coordinate_explicit}) are real. Then one can see that in Eqs.~(\ref{eq:coordinate_explicit}) and (\ref{eq:distribution_general}) $\alpha(t) > 0$ for $t > 0$. Given that $\alpha(t)\to 0$ for $t\to \infty$, function $\psi(k)$ is analytical in the upper halfplane of the complex-$k$ plane. For $q<0$ one can then add to the integral over the $k$-axis in Eq.~(\ref{eq:distribution_general}) an integral over a semicirlce $|k|\to \infty$ in the upper halfplane, which is equal to zero for $q<0$.  Since there are no singularities in the $k$-plane, the whole integral is equal to zero. Therefore once the system becomes overdamped, even in the presence of inertia $\rho(q)\equiv 0$ for $q<0$. 

A simple expression for the probability distribution can be obtained in the strongly overdamped regime, which corresponds to the limit of a large friction coefficient, $\Gamma\gg \omega_0$. In this regime 
$\lambda_1 \approx -2\Gamma$ and $\lambda_2\approx -\omega_0^2/2\Gamma$, with $|\lambda_1|\gg |\lambda_2|$; respectively, 
\[\alpha(t)\approx (2\Gamma)^{-1}\left[\exp\left(-\omega_0^2t/2\Gamma\right)-\exp(-2\Gamma t)\right].                  \]

The main contribution to $\psi(k)$ comes from the time range $t\gg 1/\Gamma$. In this range in Eq.~(\ref{eq:distribution_general}) $\alpha(t)\approx \exp(-\omega_0^2t/2\Gamma)/2\Gamma$. 
One can show that the resulting expression for $\rho(q)$  coincides with the expression for the probability distribution obtained in Ref.~\onlinecite{Dykman2010a} using a different method and in the form less convenient for the present analysis. We denote function $\psi(k)$ in this approximation as $\psi_{\rm od}^{(0)}(k)$; this function can be expressed in terms of the integral cosine  and sine, 
\begin{equation}
\label{eq:psi_overdamped_explicit}
\psi_{\rm od}^{(0)}(k)=2\frac{\Gamma^2}{\omega_0^2}\left[ \ln \frac{kg}{2\Gamma} - {\rm Ci}\left(\frac{kg}{2\Gamma}\right) - i {\rm Si}\left(\frac{kg}{2\Gamma}\right)+\gamma_{\rm E} \right]
\end{equation}
where $\gamma_{\rm E}\approx 0.58 $ is the Euler constant. As seen from this equation, parameter $k$ is scaled by the factor $\Gamma/g$. Therefore the characteristic width of the distribution (\ref{eq:distribution_general}), or in other words, the characteristic spatial scale on which the system is localized, is  $\sim g/\Gamma$.

For large $k$, from Eq.~(\ref{eq:psi_overdamped_explicit}) we have 
\begin{equation}
\label{eq:psi_overdamped_asymptotic}
\psi_{\rm od} ^{(0)}(k)\approx 2\frac{\Gamma^2}{\omega_0^2}\left[\ln\frac{kg}{2\Gamma}  + i\frac{2\Gamma}{kg}\exp\left(\frac{ikg}{2\Gamma}\right) + \gamma_{\rm E} -i\frac{\pi}{2}\right].
\end{equation}
The large dimensionless parameter used in deriving this equation is $kg/2\Gamma$.

For  $kg/2\Gamma\gg 1$ it is important also to keep in $\psi(k)$ a correction $\psi_{\rm od}^{(1)}(k)$ that comes from the time range $t\lesssim \Gamma^{-1}$ in Eq.~(\ref{eq:distribution_general}). One can find it by calculating the integral over $t$ in Eq.~(\ref{eq:distribution_general}) for large $k$ by the steepest descent method, see Sec.~\ref{sec:underdamped} where a similar but more complicated case is discussed. In the present case the integrand has one saddle point, which is located at the extremum of $\alpha(t)$ and is given by equation $\exp(-2\Gamma t)\approx (\omega_0/2\Gamma)^2$. Therefore the corresponding contribution is missed if one disregards the term $\propto\exp(-2\Gamma t)$ in $\alpha(t)$. The result reads
\begin{equation}
\label{eq:overdamped_correction}
\psi_{\rm od}^{(1)}(k)\approx -\frac{\Gamma}{\omega_0}\left(\frac{4\pi \Gamma}{kg}\right)^{1/2}
\exp\left(\frac{ikg}{2\Gamma} - i\frac{\pi}{4}\right)
\end{equation}
It is seen from this expression and Eq.~(\ref{eq:psi_overdamped_asymptotic}) that  $\psi_{\rm od}^{(1)}$ exceeds the $k$-dependent correction to the logarithmic term in $\psi_{\rm od}^{(0)}$ for large $k$.

\begin{figure}[h]
\includegraphics[width=2.8in]{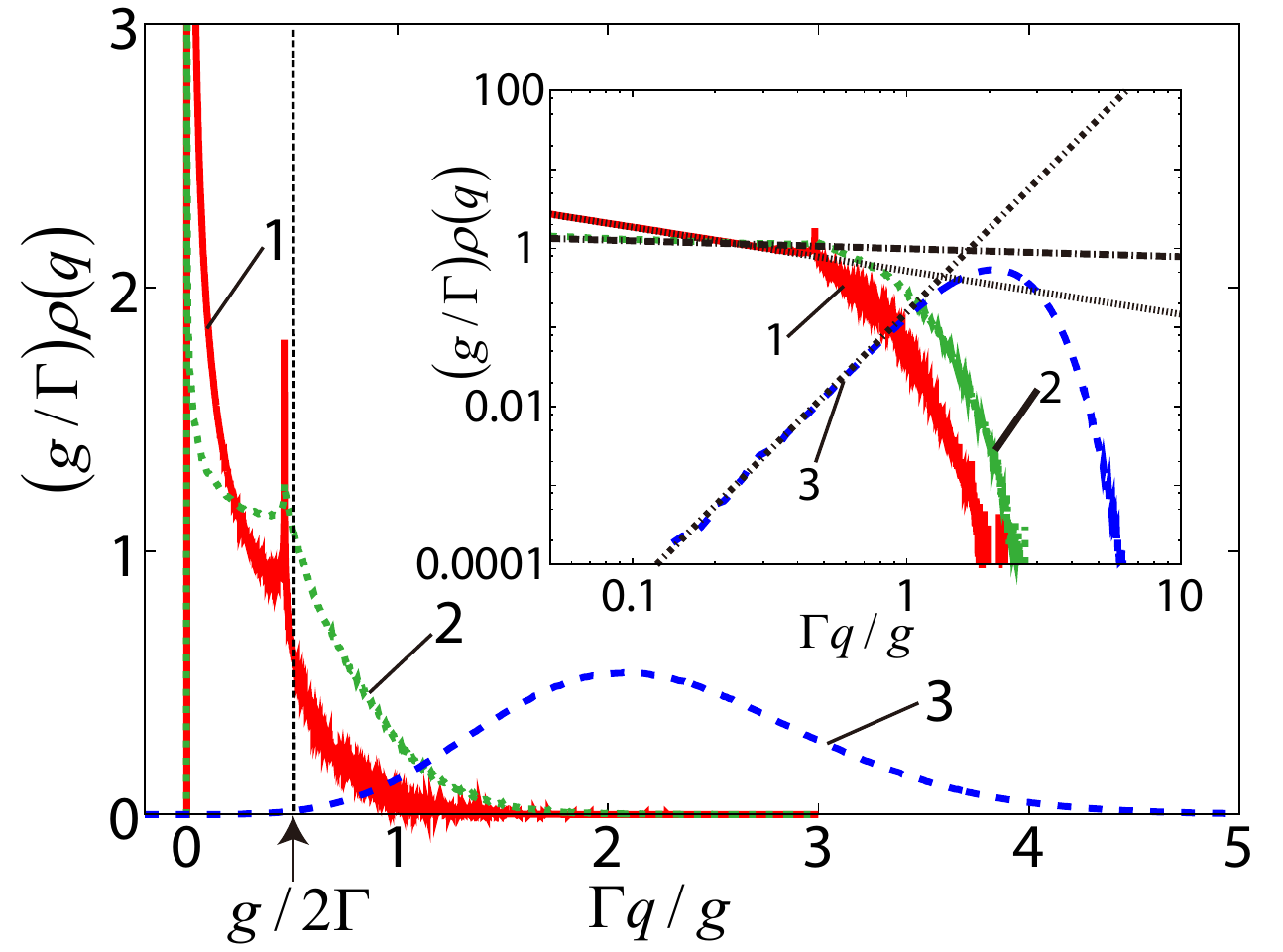}\\
\caption{The results of numerical simulations of the probability distribution of a shot-noise driven system in the overdamped regime, $\Gamma/\omega_0=3$. The data curves 1  to 3 correspond to $\nu/\Gamma = 0.025, 0.05$ and 0.25. The respective values of the critical exponent are $\beta_{\rm od}= 0.55, 0.1$ and -3.5. The inset shows the distribution on the logarithmic scale; the slopes of the straight lines are given by $-\beta_{\rm od}$.}
\label{fig:overdamped}
\end{figure}

In the integral over $k$ in Eq.~(\ref{eq:distribution_general}), we expand $\exp[-(\nu/\Gamma)\psi(k)]$ in $\psi_{\rm od}^{(1)}$ keeping the zeroth and first order terms.   As seen from Eqs.~(\ref{eq:distribution_general}) and (\ref{eq:psi_overdamped_asymptotic}), the zeroth-order term in $\psi_{\rm od}^{(1)}$ leads to a power-law behavior of $\rho(q)$ near $q=0$ for $q>0$ \cite{Dykman2010a}, 
\begin{equation}
\label{eq:scaling_overdamped}
\rho(q)\propto q^{-\beta_{\rm od}},\qquad \beta_{\rm od}=1-\frac{2\nu}{\omega_0^2/\Gamma} \qquad (\Gamma\gg\omega_0).
\end{equation}
The exponent $\beta_{\rm od}$ is determined by the ratio of the rate of Poisson pulses $\nu$ to the relaxation rate of the system $\omega_0^2/\Gamma$. For $\beta_{\rm od} <0$, i.e. for sufficiently high pulse rate, the distribution $\rho(q)$ goes to zero for $q\to +0$. On the other hand, for small pulse rate, where $\beta_{\rm od}  >  0$, the distribution diverges for $q\to 0$. This behavior is seen in Fig.~\ref{fig:overdamped}, which shows the results of numerical simulations of the equation of motion (\ref{eq:Langevin}). 

Taking into account in Eq.~(\ref{eq:distribution_general}) the term $\propto\psi_{\rm od}^{(1)}$, one sees that, for $\beta_{\rm od}> 1/2$,  the  distribution $\rho(q)$ has a peak at $q\approx g/2\Gamma$. It appears on a smooth background, and the difference $\delta\rho(q)$ from the background value is
\begin{equation}
\label{eq:overdamped_extra_singularity}
\delta\rho(q)\propto \left|q-\frac{g}{2\Gamma}\right|^{-\beta_{\rm od}+1/2}.
\end{equation}
We emphasize that the peak is asymmetric: the prefactor in $\delta\rho(q)$ depends on the sign of $q-g/2\Gamma$. The peak (\ref{eq:overdamped_extra_singularity}) does not emerge if one disregards the inertial term in the equation of motion (\ref{eq:Langevin}). For smaller $\beta_{\rm od}$ the distribution itself does not diverge for $q\to g/2\Gamma$, but for $ \beta_{\rm od} > - 1/2$ the divergence is displayed by the derivative $\partial_q \rho$.

The analysis can be extended to the case where the motion is weakly overdamped: $\Gamma > \omega_0$, but the ratio $\Gamma /\omega_0$ is not large. If, as before, $0< -\lambda_2 < -\lambda_1$, so that $\alpha(t) \propto \exp(\lambda_2 t)$ for $t\to\infty$, then to the leading order $\psi_{\rm od}^{(0)}(k)\approx (\Gamma/|\lambda_2|)\ln k$ for $k\to \infty$. The power-law singularity for $q\to +0$ is given by Eq.~(\ref{eq:scaling_overdamped}), but now $\beta_{\rm od}=1-(\nu/|\lambda_2|)$. The distribution $\rho(q)$ also has a power-law singularity at nonzero $q$. It is described by Eq.~(\ref{eq:overdamped_extra_singularity}) with the corresponding $\beta_{\rm od}$ and with the position of the singularity changed from $q=g/2\Gamma$ to $q=g\alpha(t_s)$, where $t_s$ is the root of equation $\dot \alpha(t)=0$.

The occurrence of the singular peak of the distribution for a nonzero $q$ is a benchmark of inertia in overdamped shot-noise driven systems. It is clearly seen in Fig.~\ref{fig:overdamped}. For the chosen $\Gamma/\omega_0 =3$ the position of the peak is already close to its value $g/2\Gamma$ in the limit of large $\Gamma/\omega_0$. We checked that it approaches this value with increasing $\Gamma/\omega_0$. The critical exponents of the peaks at $q=0$ and $q\approx g/2\Gamma$ are in excellent agreement with the analytical results. They approach the asymptotic values (\ref{eq:scaling_overdamped}) and (\ref{eq:overdamped_extra_singularity})  with increasing $\Gamma/\omega_0$ and become within the error of the simulations already for $\Gamma/\omega_0=10$.

\subsection{Critical regime}
\label{subsec:critical}

The results for the overdamped regime can be extended and the explicit expressions for the parameters can be obtained in the critical regime where the motion changes from overdamped to underdamped. This happens where $|\Gamma-\omega_0|\ll \Gamma$ and, consequently, $|\lambda_1 - \lambda_2|\ll |\lambda_{1,2}|\approx \Gamma$. For $\Gamma > \omega_0$ the distribution $\rho(q)=0$ for $q<0$, whereas for $\Gamma < \omega_0$ the probability to find the system in the region $q<0$ is nonzero, but $\rho(q)$  steeply decays with increasing $-q$ for small $(\omega_0-\Gamma)/\Gamma$. 

For $\Gamma=\omega_0$ we have $\alpha(t)=t\exp(-\Gamma t)\geq 0$, and therefore still $\rho(q)=0$ for $q<0$. Using the arguments that led to Eqs.~(\ref{eq:scaling_overdamped}) and (\ref{eq:overdamped_extra_singularity}) one obtains that the power-law singularity of $\rho(q)$ for $q\to +0$ has the form $q^{-\beta_{\rm cr}}$ with $\beta_{\rm cr}=1- (\nu/\Gamma)$. The distribution $\rho(q)$ also has a power-law singularity of the type of Eq.~(\ref{eq:overdamped_extra_singularity}), which is located at $q=g/e\Gamma$ and is characterized by exponent $\beta_{\rm cr}-1/2$. 

For small $(\omega_0-\Gamma)/\Gamma >0$ there emerge additional power-law singularities of $\rho(q)$. However, they are located at exponentially small $|q|\propto \exp[-\pi\Gamma/(\omega_0-\Gamma)]$ and therefore are extremely hard to resolve.

\section{Underdamped regime}
\label{sec:underdamped}

The general expression for the probability distribution simplifies also in the case of small relaxation rate, $\Gamma \ll \omega_0$. In this case the motion of the system in the absence of noise is weakly damped vibrations at frequency $\approx \omega_0$, and $\lambda_{1,2}\approx -\Gamma\pm i\omega_0$. Noise pulses excite vibrations at random. Clearly, the stationary probability distribution $\rho(q)$ in the presence of dissipation and noise is no longer limited to the region $q\geq 0$, it is expected to be almost symmetric  with respect to $q$. 

To describe $\rho(q)$ we note that in Eq.~(\ref{eq:coordinate_explicit}) 
\[\alpha(t)\approx \omega_0^{-1}\exp(-\Gamma t)\sin\omega_0t.\] 
For an underdamped system the function $\psi(k)\equiv \psi_{\rm ud}(k)$  in Eq.~(\ref{eq:distribution_general}) to the leading order in $\Gamma/\omega_0$ has the form
\begin{eqnarray}
\label{eq:psi_underdamped_smooth}
\psi_{\rm ud}(k)&\approx& \psi_{\rm ud}^{(0)}(k) =\int\nolimits_0^{kg/\omega_0}dx [1 -J_0(x)]/x.
\end{eqnarray}
Function $\psi_{\rm ud}^{(0)}$ describes the smooth part of the distribution $\rho(q)$. The typical width of the distribution, as seen from Eqs.~(\ref{eq:distribution_general}) and (\ref{eq:psi_underdamped_smooth}), is $\sim g/\omega_0$.

In the range $kg/\omega_0\gg 1$ we have
\begin{eqnarray}
\label{eq:asymptotic_ud}
&&\psi_{\rm ud}^{(0)}(k)\approx \ln\frac{kg}{2\omega_0} + \gamma_{\rm E},
\end{eqnarray}
which indicates that, even in the underdamped case, the distribution has a power-law singularity for $|q|\to 0$. Generally, this singularity appears on a smooth background. From Eqs.~(\ref{eq:distribution_general}) and (\ref{eq:asymptotic_ud}) the difference $\delta\rho(q)$ from the background value for small $|q|$ is
\begin{equation}
\label{eq:q_origin_underdamped}
\delta\rho(q) \propto |q|^{-\beta_{ud}}, \qquad \beta_{\rm ud}= 1 - \frac{\nu}{\Gamma}\qquad (\Gamma\ll \omega_0).
\end{equation}
In contrast to the overdamped case, Eq.~(\ref{eq:scaling_overdamped}), the singular behavior occurs on the both sides of $q=0$, with the same exponent $\beta_{\rm ud}$. This exponent is again determined by the ratio of the pulse rate $\nu$ to the relaxation rate, which is equal to $\Gamma$ for weak damping. 

The approximation leading to Eq.~(\ref{eq:q_origin_underdamped}) is justified for $\beta_{\rm ud} > - 2$. In the opposite case, $\beta_{\rm ud}  < - 2$, the distribution for small $q$ is formed primarily by the region where $kg/\omega_0 \lesssim 1$ and the distribution is parabolic near $q=0$. Still there may be singularities in the derivatives of $\rho$ of sufficiently high order; in what follows we assume $\beta_{\rm ud} > - 2$.

\subsection{Multiple distribution peaks away from the origin}
\label{subsec: multiple_coordinate}

\begin{figure}[h]
\includegraphics[width=1.5in,angle=90]{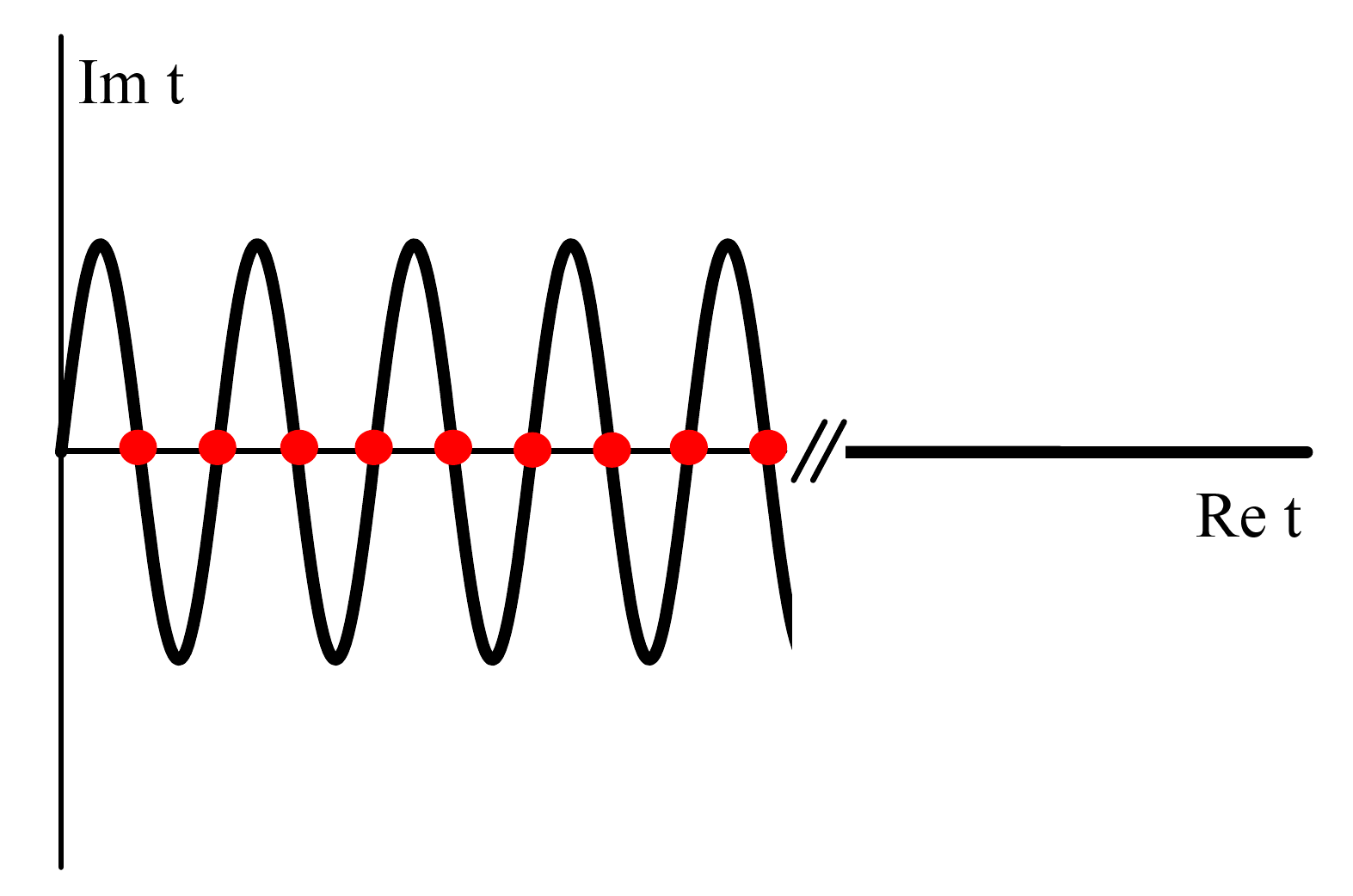}\\
\caption{The contour of integration over time for calculating function $\psi(k)$, Eq.~(\ref{eq:distribution_general}), for an underdamped system for $kg/\omega_0\gg 1$. In the range   $\exp(-\Gamma t')kg/\omega_0 \gg 1$ ($t'\equiv {\rm Re}~t$) the contour is oscillating and goes through the saddle points $t_n\approx (n+1/2)\pi/\omega_0$ of the function $\exp[ikg\alpha(t)]$, making angles $(-1)^{n+1}\pi/4$ with the Re~$t$-axis ($n=0,1,2,\ldots$). The saddle points are indicated by the solid circles. For large Re~$t$ the integration contour approaches the Re~$t$-axis. In the overdamped limit the integration contour used to obtain Eq.~(\ref{eq:overdamped_correction}) crosses the Re~$t$-axis only once for $\exp(-\Gamma t)=\omega_0/2\Gamma$ and then approaches the Re~$t$-axis.  }
\label{fig:integration_contour}
\end{figure}

The overall distribution $\rho(q)$ in the underdamped case turns out to have multiple singularities. In their analysis one should take into account the terms $\sim \Gamma/\omega_0$ in $\psi_{\rm ud}(k)$. In the limit of large $k$ they can be found by calculating the integral over time in Eq.~(\ref{eq:distribution_general}) by the steepest descent method. It requires bending the contour of integration over time, which is justified since $\alpha(t)$ has no singularities near the Re~$t$-axis, see Fig.~\ref{fig:integration_contour}.

As seen from Fig.~\ref{fig:integration_contour}, for large $k$ the function $\psi_{\rm ud}(k)$ has two major contributions. One comes from the region of large time, where $kg\alpha(t)\lesssim 1$ and the integration goes along the Re~$t$-axis. To the leading order in $k$, it is given by the logarithmic term in $\psi_{\rm ud}^{(0)}(k)$, Eq.~(\ref{eq:asymptotic_ud}). It is not proportional to $\Gamma/\omega_0\ll 1$. The other contribution comes from smaller times, where the exponential term $\exp[ikg\alpha(t)]$  in $\psi(k)$ in Eq.~(\ref{eq:distribution_general}) can significantly differ from 1. In this region, if the integration contour is appropriately bent, this term has multiple saddle points where $\dot \alpha(t)=0$. These points are marked in Fig.~\ref{fig:integration_contour}. The resulting contribution is $\propto \Gamma/\omega_0$ and has the form
\begin{eqnarray}
\label{eq:psi1_underdamped}
\psi_{\rm ud}^{(1)}(k) &\approx&  -\frac{\Gamma}{\omega_0}\left(\frac{2\pi\omega_0}{kg}\right)^{1/2}\,  \sum_{n=0}^{n_{\max}} e^{\Gamma t_n/2}\exp[i\phi_n(k)]\nonumber\\
\phi_n(k) &= & (-1)^n\frac{kg}{\omega_0}e^{-\Gamma t_n} -(-1)^n\frac{\pi}{4},\\
 t_n &=&\pi\omega_0^{-1} (n + 1/2).\nonumber
\end{eqnarray}
Here, parameter $n_{\max}$ is determined by the condition $(kg/\omega_0)\exp(-\Gamma t_n) \gg 1$ for $n < n_{\max}$. Therefore $|\psi_{\rm ud}^{(1)}|\ll 1$, and in Eq.~(\ref{eq:distribution_general})  $\exp[-(\nu/\Gamma)\psi_{\rm ud}^{(1)}]$ can be expanded in $\psi_{\rm ud}^{(1)}$; we will keep the first-order term in this expansion. 

From Eqs.~(\ref{eq:distribution_general}) and (\ref{eq:psi1_underdamped}), for sufficiently low rate of noise pulses $\nu/\Gamma$, where $\beta_{\rm ud}  > 1/2$,  distribution $\rho(q)$ has multiple power-law peaks. The deviation of $\rho(q)$ from the smooth background near the $n$th peak $\delta\rho_n(q)$ is
\begin{equation}
\label{eq:scaling_underdamped}
\delta\rho_n(q) \propto |q-q_n|^{-\beta_{\rm ud} + 1/2}, \qquad q_n=(-1)^n\frac{g}{\omega_0}e^{-\Gamma t_n}.
\end{equation}
As seen from Eqs.~(\ref{eq:psi1_underdamped}) and (\ref{eq:scaling_underdamped}), the positions $q_n$ of the singular peaks of $\rho(q)$ form a geometric progression. All peaks display a power-law shape with the same exponent $\beta_{\rm ud} - 1/2$. At the same time, the prefactor in $\delta\rho_n(q)$ takes on different values on the opposite sides of the peak, i.e., it depends on the sign of $q-q_n$. Equations ~(\ref{eq:psi1_underdamped}) and (\ref{eq:scaling_underdamped}) make it possible to find the prefactor (this can also be done in the overdamped regime), but the expression is somewhat cumbersome

\begin{figure}[h]
\includegraphics[width=2.8in]{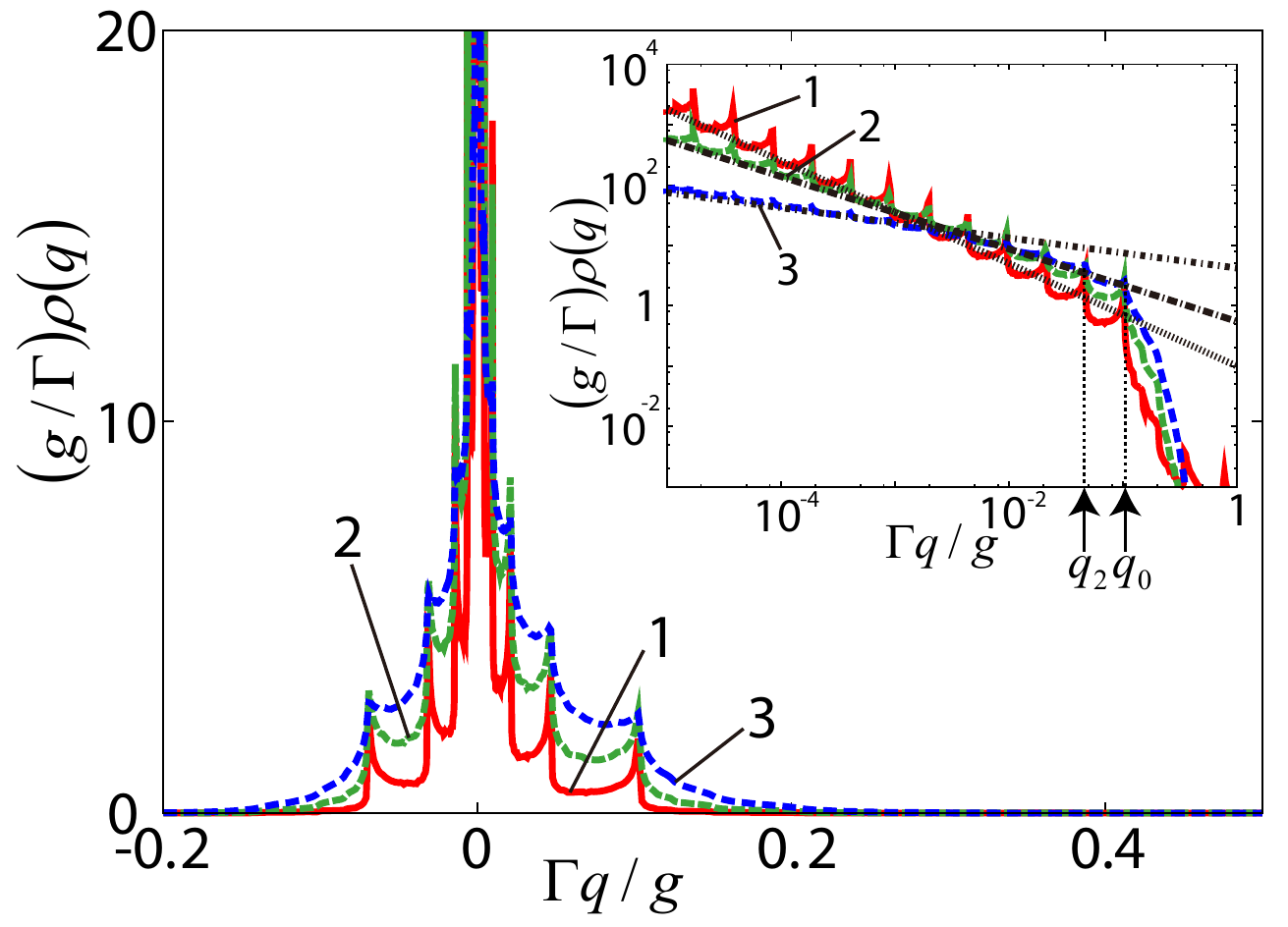}\\
\caption{The results of numerical simulations of the probability distribution of a shot-noise driven system in the underdamped regime, $\Gamma/\omega_0=0.125$. The data curves 1 to 3 correspond to $\nu/\Gamma = 0.15, 0.4$, and 0.75. The respective values of the critical exponent are $\beta_{\rm ud}= 0.85, 0.6$ and 0.25. The inset shows the distribution on the logarithmic scale; the slopes of the straight lines are given by $\beta_{\rm ud}$.}
\label{fig:underdamped}
\end{figure}

\begin{figure}[h]
\includegraphics[width=3.2in]{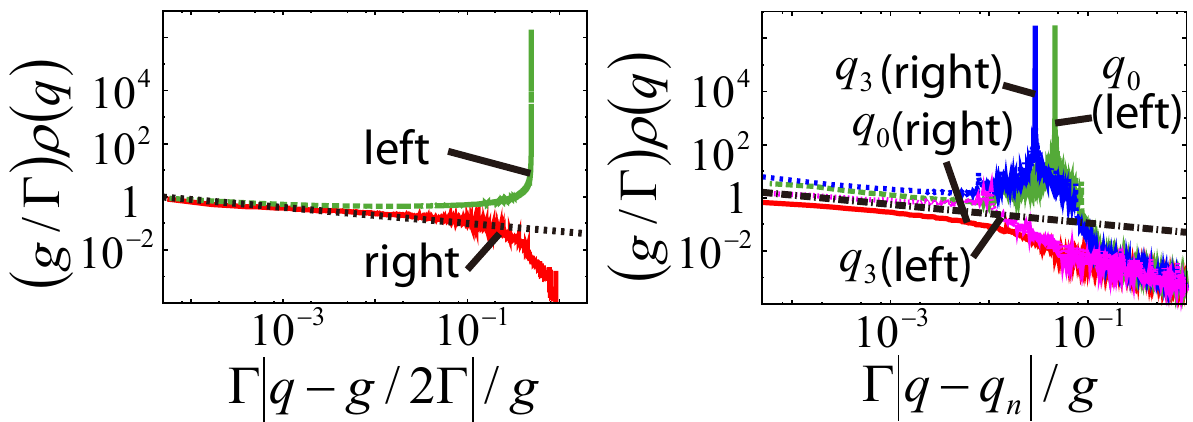}\\
\caption{The vicinities of the peaks of the distribution $\rho(q)$ that lie away from $q=0$ for the overdamped (left panel) and underdamped (right panel) systems. The distances are counted off from the positions of the corresponding peaks $q_n$; in the overdamped case there is only one such peak, whereas for the underdamped case we chose peaks with $n=0$ and 3, Eq.~(\protect\ref{eq:scaling_underdamped}). The labels ``right" and ``left" indicate the right and left side of the peak, i.e.,  $q-q_n>0$ and $q-q_n<0$, respectively.  In the left panel $\Gamma/\omega_0=10, \, \beta_{\rm od}=0.8$; in the right panel $\Gamma/\omega_0=0.05, \, \beta_{\rm ud}= 0.85$.  The straight lines show the expected asymptotic  slopes of $\ln\rho(q)$ for $q\to q_n$. The maxima for comparatively large $|q-q_n|$ visible in the right panel correspond to other peaks of $\rho(q)$, i.e., the peaks with $n\neq 0,3$.}
\label{fig:q_exponents}
\end{figure}

The peaks with largest $|q_n|$ are the ones with $n=0$ and $n=1$. They lie on the opposite sides of the center of the distribution at $q=0$. The positions $q_0$ and $q_1$ are asymmetric for nonzero $\Gamma/\omega_0$, but the asymmetry is weak for small $\pi\Gamma/\omega_0$. Other peaks lie between $q_0$ and $q_1$. Their amplitudes increase with decreasing $|q_n|$ because of the factor $\exp(\Gamma t_n/2)$. We note again that these amplitudes are $\propto \Gamma/\omega_0\ll 1$. For higher rates of shot-noise pulses, where $1/2 > \beta > -1/2$, the peaks of $\rho(q)$ disappear in the asymptotic theory (\ref{eq:scaling_underdamped}), but a self-similar structure of power-law divergences  can still be seen in the derivative of the distribution $\partial_q\rho$.

The predicted behavior is in agreement with the results of numerical simulations of Eq.~(\ref{eq:Langevin}) shown in Fig.~\ref{fig:underdamped}. The distribution obtained by simulations displays power-law singularities, and the positions of these singularities are in excellent agreement with Eq.~(\ref{eq:scaling_underdamped}). The exponent of the singularity for $q=0$ is also in agreement with the theory. For the peaks of $\rho(q)$ at $q\neq 0$ in the case of the smallest $\nu/\Gamma$, where $\beta_{\rm ud} = 0.85$ and these peaks are most pronounced, the exponents agree with the theory. However, for the moderately small $\Gamma/\omega_0$ used in Fig.~\ref{fig:underdamped}, for larger $\nu/\Gamma$ the agreement is worse: for $\beta_{\rm ud}=  0.6$ instead of the expected exponent $0.1$, see Eq.~(\ref{eq:scaling_underdamped}), the obtained exponent is $\sim 0.16$, which we believe is due to the overlapping of the peaks that complicates retrieving the exponent.

Careful studies for smaller $\Gamma/\omega_0$ demonstrated an excellent agreement of both the positions of the peaks and the exponents with the theory. The exponents are the same on the both sides of the peaks and are the same for all peaks at $q_n\neq 0$. The results are shown in Fig.~\ref{fig:q_exponents}. This figure also demonstrates an excellent agreement with the theory of the exponents obtained by numerical simulations for an overdamped system.

An important feature of the singularities of the distribution $\rho(q)$ is that their {\it positions} $q_n$ depend only on the area of the noise pulses $g$ and the system parameters, but not on the pulse rate $\nu$. In contrast, the exponent $\beta_{\rm ud}$ depends on $\nu$ scaled by the system relaxation rate, but is independent of the noise pulse area $g$.

\section{Singular peaks in the energy probability distribution}
\label{sec:energy}

For a Poisson-noise driven system the probability distribution over the system energy $\rho_E(E)$ differs from the Boltzmann distribution. The tail of this distribution in the limit of small damping was discussed earlier \cite{Sukhorukov2009}. Here we also consider the small damping case, $\Gamma\ll \omega_0$, but we are interested in the central part of the distribution. We show that the distribution is singular and can have multiple power-law peaks. 

Function $\rho_E(E)$ can be conveniently expressed in terms of the distribution of the system in phase space $\rho_{q,p}(q,p)$. For a linear system described by the equation of motion (\ref{eq:Langevin}) the stationary distribution $\rho_{q,p} = \langle \delta[q-q(t)]\delta[p-p(t)]\rangle$  can be found using the explicit expression (\ref{eq:coordinate_explicit}) for the system coordinate $q(t)$ and the corresponding expression for $p(t)=\dot q(t)=\int\nolimits_{-\infty}^tdt' f_P(t')\partial_t\alpha(t-t')$,
\begin{widetext}
\begin{eqnarray}
\label{eq:rho_E_defined}
&&\rho_E(E)=\int dq\,dp \rho_{q,p}(q,p)\delta[E-E(q,p)],\qquad E(q,p)=\left(p^2 + \omega_0^2q^2\right)/2,\nonumber\\
&&\rho_{q,p}(q,p)=(2\pi)^{-2}\int dk_q\,dk_p\exp\left[-i(k_qq+k_pp) - \frac{\nu}{\Gamma}\psi_E({\bf k})\right] \qquad [{\bf k}=(k_q,k_p)],\\
&&\psi_E({\bf k})=\Gamma\int\nolimits_0^{\infty}dt\left\{1-\exp\left[igf_{\bf k}(t)\right]\right\},\qquad
f_{\bf k}(t)=k_q\alpha(t) + k_p\dot\alpha(t). \nonumber
\end{eqnarray}
\end{widetext}
This expression is a straightforward extension of Eq.~(\ref{eq:distribution_general}). 

The singular behavior of the distribution $\rho_E$ is determined by the behavior of the function $\psi_E$ in the range of large $|k_q|, |k_p|$. The leading-order contribution $\psi_E^{(0)}$ to $\psi_E$ comes from the region of $t$ in Eq.~(\ref{eq:rho_E_defined}) where the term $\exp[igf_{\bf k}(t)]$ is fast oscillating and can be disregarded. The size of this region can be estimated by noticing that $\alpha(t),\dot\alpha(t) \propto \exp(-\Gamma t)$. Therefore
\begin{equation}
\label{eq:psi_qp_leading}
\psi_E^{(0)}\approx \ln\left[g \max (|k_q|\omega_0^{-1}, |k_p|)\right].
\end{equation}

From Eqs.~(\ref{eq:rho_E_defined}) and (\ref{eq:psi_qp_leading}), the distribution $\rho_E$ has a power-law singularity at $E=0$, 
\begin{equation}
\label{eq:rho_E_for_E_to_0}
\rho_E(E)\propto E^{-\beta_E},\qquad \beta_E= 1 - \frac{\nu}{2\Gamma} \qquad (E\to +0).
\end{equation}
This singularity corresponds to a divergent (but integrable) peak for $\beta_E > 0$; if $\beta _E > -1$ the derivative $\partial_E\rho_E$ diverges for $E\to +0$.

It is interesting to compare the singularity (\ref{eq:rho_E_for_E_to_0}) with the singularity of the coordinate distribution $\rho(q)$ of the underdamped system for $q\to 0$, Eq.~(\ref{eq:q_origin_underdamped}). The exponent $\beta_E$ is expressed in terms of the exponent of $\rho (q)$ as $\beta_E=(\beta_{\rm ud} + 1)/2$. Therefore, if $\beta_{\rm ud} > 0$ and $\rho(q)$ has a peak for $q= 0$, the energy distribution also has a peak for $E\to +0$. However, the energy distribution can have a peak for $E= +0$ even where $\beta_{\rm ud} < 0$ and $\rho(q)$ does not have a peak for $q=0$.

\subsection{Multiple peaks of the energy distribution for nonzero energy}
\label{subsec:nonzero_energy}

The  contribution $\psi_E^{(1)}$ to $\psi_E$ of the subleading order in $k_q, k_p$ leads to the onset of singularities in the distribution $\rho_E$ for $E>0$. To find them we first change to cylindrical coordinates in ${\bf k}$-space,
\[\omega_0^{-1}k_q = \kappa \cos\theta, \quad k_p=\kappa\sin\theta.\]
Then, changing in Eq.~(\ref{eq:rho_E_defined}) from $q,p$ to the standard action-angle variables \cite{LL_Mechanics2004} and integrating over the angle, we obtain
\begin{equation}
\label{eq:cylindrical}
\rho_E(E)= (2\pi)^{-1}\int \kappa\,d\kappa\,d\theta J_0(\kappa\sqrt{2E})e^{-(\nu/\Gamma)\psi_E}.
\end{equation}

In variables $(\kappa,\theta)$, to first order in $\Gamma/\omega_0$ function $f_{\bf k}$ in the integrand of $\psi_E$ becomes
\begin{eqnarray}
\label{eq:exponent_E}
f_{\bf k}(t)&\approx& \kappa \exp(-\Gamma t)
\nonumber\\   
&&\times
\left[\sin(\omega_0t+\theta) -(\Gamma/\omega_0)\sin\theta \sin\omega_0t\right].
\end{eqnarray}
It is seen from this equation that, for $\kappa \to \infty$, the leading-order term in $\psi_E$ is $\psi_E^{(0)}\approx \ln (\kappa g)$, which coincides with  Eq.~(\ref{eq:psi_qp_leading}).  We note, however, that variables $(\kappa,\theta)$ are less convenient for calculating the behavior of $\rho_E$ for $E\to 0$ than $(k_q,k_p)$ because of the singular nature of the integrals; nevertheless, the above expression for $\psi_E^{(0)}$ and Eq.~(\ref{eq:cylindrical}) immediately show that the scaling of $\rho_E$ for $E\to +0$  is indeed of the form of Eq.~(\ref{eq:rho_E_for_E_to_0}).

The subleading term $\psi_E^{(1)}$ can be obtained by calculating the integral over time in Eq.~(\ref{eq:rho_E_defined}) by the steepest descent method, as in Sec.~\ref{sec:underdamped}, cf. Fig.~\ref{fig:integration_contour}. From Eq.~(\ref{eq:exponent_E}), to the lowest order in $\Gamma/\omega_0$ the saddle points are 
\begin{equation}
\label{eq:t_n_theta}
t_n(\theta)=\pi\omega_0^{-1}(n+1/2)-\omega_0^{-1}\theta
\end{equation}
[the instants $t_n$ in Eq.~(\ref{eq:psi1_underdamped}) are equal to $t_n(0)$].

The result of the integration has the form similar to Eq.~(\ref{eq:psi1_underdamped}),
\begin{eqnarray}
\label{eq:psi_E_1}
&&\psi_E^{(1)}\approx - (\Gamma/\omega_0)\left(2\pi/\kappa g\right)^{1/2}
\sum_{n}e^{\Gamma t_n(\theta)/2}a_n(\kappa,\theta),\nonumber\\
&&a_n(\kappa,\theta)=
\exp\left[igf_{\bf k}\bigl(t_n(\theta)\bigr) - (-1)^ni\pi/4\right].
\end{eqnarray}
Here we have disregarded corrections $\propto \Gamma/\omega_0$ unless they are multiplied by a large factor. In particular, we keep the term $\propto \Gamma/\omega_0$ in function $f_{\bf k}$, since it is multiplied by $\kappa$. It can be large for large $\kappa$ and such $t_n(\theta)$ that $\kappa g\exp[-\Gamma t_n(\theta)] > \omega_0/\Gamma\gg 1$.  To first order in $\Gamma/\omega_0$
\[
f_{\bf k}\bigl(t_n(\theta)\bigr) \approx \kappa e^{-\Gamma t_n(0)}(-1)^n
\left[1 +\frac{\Gamma}{\omega_0}\left(\theta - \frac{1}{2}\sin 2\theta\right)\right] .
\]
The condition $\kappa g\exp[-\Gamma t_n(\theta)] \gg 1$ imposes the upper limit on $n$ in the sum over $n$ in Eq.~(\ref{eq:psi_E_1}).

To find the most pronounced singularities of $\rho_E$ we expand in Eq.~(\ref{eq:cylindrical}) $\exp[-(\nu/\Gamma)\psi_E^{(1)}]$ to the first order in $\psi_E^{(1)}$. Then the calculation of $\rho_E$ reduces to integrating $\psi_E^{(1)}$ over $\theta$, which has to be followed by integration over $\kappa$ with the appropriate weight. Since from Eq.~(\ref{eq:rho_E_defined}) $t_n(\theta)\geq 0$, it is convenient to integrate over $\theta$ from $-3\pi/2$ to $\pi/2$, which is seen from Eq.~(\ref{eq:t_n_theta}) to correspond to $n\geq 0$ in Eq.~(\ref{eq:psi_E_1}).

For large $\kappa$, integration of $\exp[igf_{\bf k}\bigl(t_n(\theta)\bigr)]$  over $\theta$ can be done by the stationary phase method. The stationary points, $\partial_{\theta}f_{\bf k}=0$, are located at $\theta_{st}= m\pi$ with integer $m$. At these points $\partial_{\theta}^2 f_{\bf k}=0$. Therefore 
\begin{equation}
\label{eq:int_a_n}
\int d\theta a_n(\kappa,\theta) \approx \frac{C_n}{(\kappa g)^{1/3}}\left[a_n(\kappa,0) + a_{n+1}^*(\kappa,0)\right],
\end{equation}
where 
\[C_n=\left\{(2\Gamma/3\omega_0)(-1)^n e^{-\Gamma t_n(0)}\right\}^{-1/3}\Gamma(1/3)/\sqrt{3}\]
($\Gamma(x)$ is the Gamma function).

Taking into account that $J_0(x)\propto x^{-1/2}\cos(x-\pi/4)$ for $x \gg 1$, one obtains from Eqs.~(\ref{eq:cylindrical}) --  (\ref{eq:int_a_n}) that $\rho_E(E)$ displays power-law singularities for nonzero $E$, and near an $n$th singularity the deviation $\delta\rho_{E,n}(E)$ of $\rho_E$ from the smooth background is
\begin{eqnarray}
\label{eq:singular_E_n}
&&\delta\rho_{E,n}(E)\propto |E-E_n|^{-2\beta_E + 4/3},\nonumber\\
&&E_n=\frac{1}{2}g^2\exp[-2\Gamma t_n(0)].
\end{eqnarray}
As in the case of the singularities of the coordinate distribution for an underdamped system, for all singularities of $\rho_E$ the exponents are the same  and the positions of the singularities form a geometric progression. The exponent $2\beta_E - 4/3 = (2/3)-(\nu/\Gamma)$ is independent of the area of the noise pulses $g$, whereas the positions of the singularities $E_n$ are independent of the pulse rate $\nu$. In the most interesting case of comparatively low pulse rate, where $2\beta_E - 4/3 > 0$, the singularities correspond to the peaks of the distribution. We note that the condition $2\beta_E - 4/3 > 0$ holds for higher pulse rate $\nu$ than the condition $\beta_{\rm ud} - 1/2 > 0$, which is necessary for observing multiple peaks of the coordinate distribution.
\begin{figure}[h]
\includegraphics[width=2.8in]{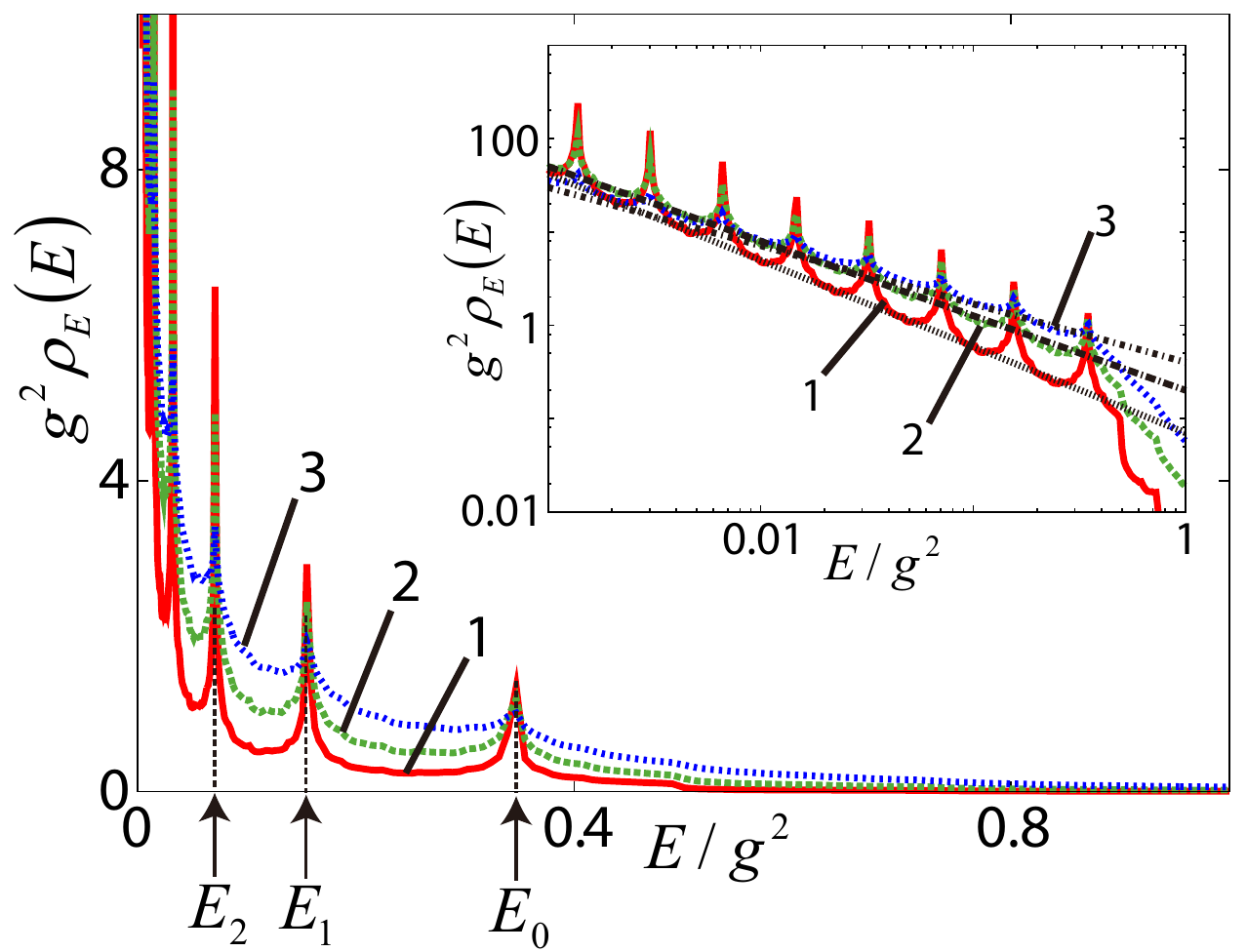}\\
\caption{The results of numerical simulations of the probability distribution over the energy of a shot-noise driven underdamped system, where $\Gamma/\omega_0=0.125$. The data curves 1 to 3 correspond to $\nu/\Gamma = 0.15, 0.4$ and 0.75. The respective values of the critical exponent for $E\to 0+$, which are shown by straight lines in the inset, are $\beta_{E}= 0.925, 0.8$ and 0.625. }
\label{fig:energy}
\end{figure}

In Fig.~\ref{fig:energy} we compare the predictions with the results of numerical simulations. The positions of the peaks of $\rho_E$ are in excellent agreement with Eq.~(\ref{eq:singular_E_n}) already for moderately small $\Gamma/\omega_0 = 1/8$. As expected, the peaks become less pronounced with the increasing noise pulse rate. The singularities of $\rho_E$ for $E=E_n>0$ are still visible even for $2\beta_E - 4/3 < 0$ (data curve 3 in Fig.~\ref{fig:energy}); the asymptotic theory predicts that, for the corresponding $\beta_E$, the derivative $\partial_E\rho_E$ should diverge for $E=E_n$. The exponent of the peak for $E= +0$ obtained numerically is also in excellent agreement with the theory.

\begin{figure}[ht]
\includegraphics[width=2.8in]{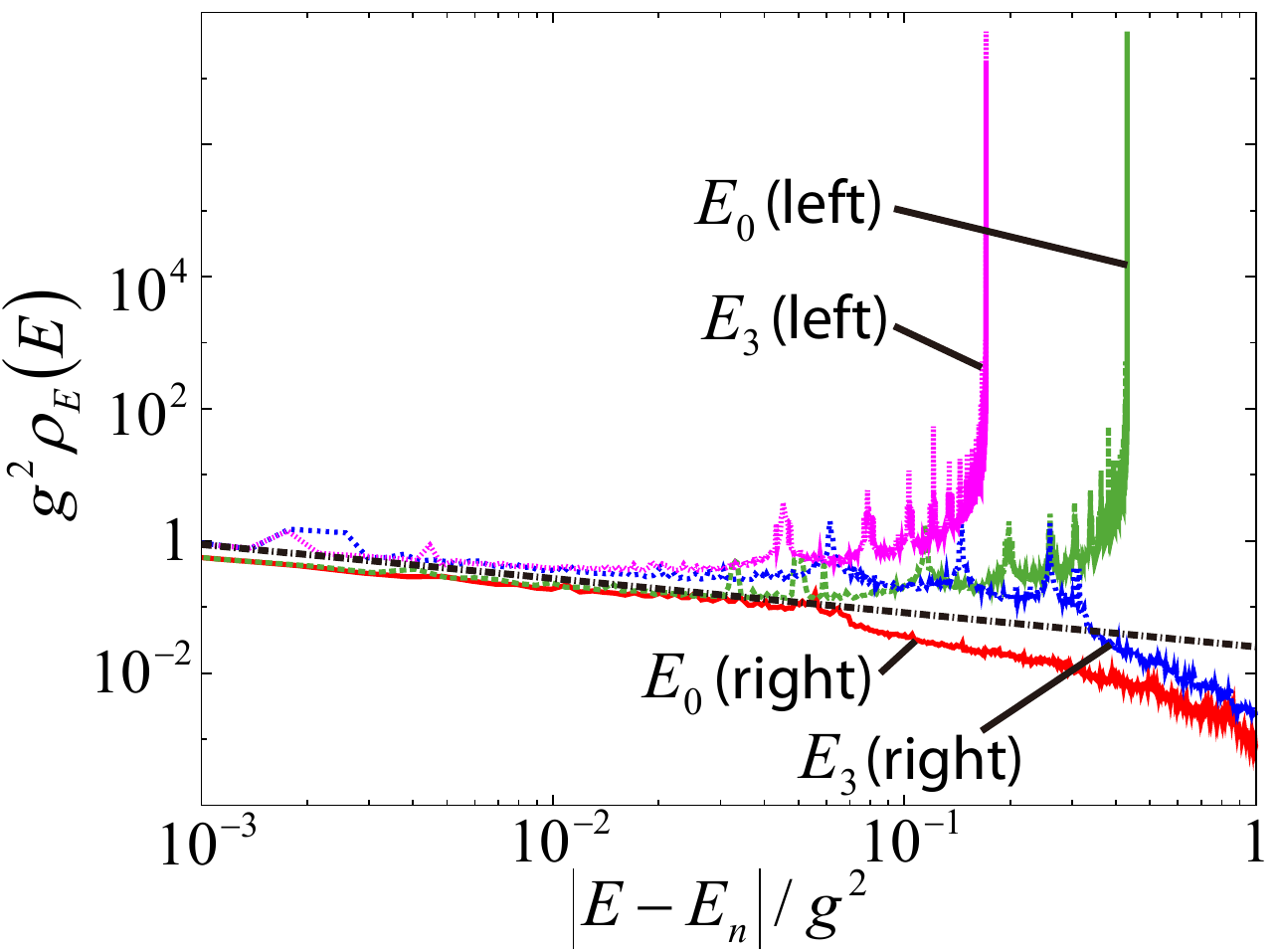}\\
\caption{The results of numerical simulations of the probability distribution over energy of an underdamped system near the peaks (\protect\ref{eq:singular_E_n}) with $n=0$ and 3. The labels ``right" and ``left" refer to the right and left sides of the peaks, where $E-E_n>0$ and $E-E_n<0$, respectively. The parameters are $\Gamma/\omega_0=1/20, \beta_E = 0.925$ ($\nu/\Gamma=0.15$). The straight line shows the analytical value of the exponent of the power-law peaks for $E\to E_n$. }
\label{fig:energy_exponents}
\end{figure}

In Fig.~\ref{fig:energy_exponents} we present the results of simulations of the singularities very close to the peaks of $\rho_E$. We find that the singularities are well described by the power law, and the exponents are in excellent agreement with the analytical theory. It should be noted that the peaks found in numerical simulations had a structure of doublets, with extremely small distance between the peaks in the doublet. Analytically one might expect a doublet structure for not too small $\Gamma/\omega_0$, as seen from Eq.~(\ref{eq:int_a_n}). However, the corresponding analysis is beyond the accuracy of the asymptotic theory developed here.

\section{Qualitative picture of the onset of the distribution singularities}
\label{sec:discussion}

The onset of singularities of the probability distribution of  Poisson-noise driven systems can be understood by noticing that a single noise pulse shifts the momentum of the system by $g$. If the pulse rate $\nu$ is small compared to the relaxation rate $t_r^{-1}$, on average the system has time to relax between the pulses and to approach the equilibrium position $q=p=0$. This leads to accumulation of the probability distribution near the equilibrium position for $\nu t_r \ll 1$.  As we showed, the result is a power-law peak of $\rho(q)$ at $q=0$ with exponent $1 - \nu t_r$. 

The argument and the result apply to both overdamped and  underdamped systems; for strongly overdamped systems $t_r= 2\Gamma/\omega_0^2$, whereas for strongly underdamped systems $t_r=1/\Gamma$. The expression $1-\nu t_r$  for the exponent  agrees with the qualitative picture that, for $\nu\to 0$, where the pulses are rare, the distribution should be proportional to the reciprocal velocity for a given $q$, which is itself proportional to $q$ (or to $|q|$, in the underdamped case).  On the other hand, for large $\nu$ the noise becomes effectively Gaussian and the distribution becomes smooth near the maximum, see below.  We note that even in the large-$\nu t_r$ limit $\rho(q)$ has a power-law cutoff for $gq\to +0$ in the case of an overdamped system. 

The power-law singularity emerges also in the critical regime where the system dynamics changes from overdamped to underdamped. A similar singularity, and with the same exponent, can be shown to characterize the distribution over the momentum $p$.
Interestingly, power-law singularities of the distribution emerge also in overdamped systems driven by another important type of non-Gaussian noise, the telegraph noise \cite{Klyatskin1977,*Kitahara1979,*Horsthemke1984}.

An underdamped system excited by a single noise pulse performs weakly damped vibrations. If initially the system was at the  equilibrium position, the momentum right after the pulse is $p=g$, and the ensuing vibrations in the weak-damping limit have the form 
\[q(t)\approx (g/\omega_0)e^{-\Gamma t} \sin\omega_0t \qquad (\omega_0\gg \Gamma);\]
here, we count time off from the instant when the pulse occurred.

The slowing down at the turning points $\dot q = p =0$ leads to the peaks of the probability distribution. From the above expression, such peaks should be located at $q=q_n=(-1)^n(g/\omega_0)\exp(-\Gamma t_n)$ with $t_n=\pi\omega_0(n+1/2)$, in agreement with Eqns.~(\ref{eq:psi1_underdamped}) and (\ref{eq:scaling_underdamped}). As we showed, they are described by a power-law with the exponent $(1/2) - \nu t_r$. This exponent differs from the exponent of the peak at $q=0$. The onset of the peaks at $q_n\neq 0$ requires a lower value of $\nu t_r$ than for the $q=0$-peak. 

The fact that the exponent approaches $1/2$ for $\nu\to 0$ can be again understood as a result of the probability distribution being proportional to the reciprocal velocity in this limit. Near the extrema of $q(t)$ the velocity scales as $|q-q_n|^{1/2}$. We note that the peaks of $\rho(q)$ become strongly asymmetric for small $\nu t_r$, in agreement with the above argument.

In the case of an overdamped system, after the pulse-induced increase of the momentum $p=0\to p=g$, the system coordinate first moves away from equilibrium and then monotonically comes back,
\[ q(t) \approx (g/2\Gamma)\left(e^{-\omega_0^2t/2\Gamma} - e^{-2\Gamma t}\right) \qquad (\omega_0\ll \Gamma).\]
Respectively, the probability distribution over $q$ has a peak where $q(t)$ is maximal, $q\approx g/2\Gamma$. We showed that the peak is described by a power-law with an exponent $(1/2) - \nu t_r$ for $\nu t_r < 1/2$.  

Similar arguments can be applied to the peaks of the energy distribution in an underdamped system $\rho_E$. This distribution can have a power-law peak for $E\to 0$ due to the accumulation of the probability density near $q=p=0$ for rare noise pulses. To understand the peaks for nonzero energies we note that, for low pulse rate, after a noise pulse at $t=0$ the energy evolves as 
\[ E(t)\approx \frac{1}{2}g^2\exp(-2\Gamma t)\left[ 1 - (\Gamma/\omega_0)\sin 2\omega_0t\right]\]
to first order in $\Gamma/\omega_0$. Function $E(t)$ has inflection points at $t_n=\pi\omega_0^{-1}(n+1/2)$. The peaks of $\rho_E$ occur as a consequence of the slowing down near these points. The values $E_n = E(t_n)$ give the positions of the peaks, in agreement with Eq.~(\ref{eq:singular_E_n}). The corresponding exponent is $(2/3) - \nu t_r$, the same for all peaks. The limiting value of the exponent for $\nu\to 0$ can be again understood by noticing that $\rho_E$ in this case is given by the reciprocal rate of the change of $E(t)$ near the inflection point.

In real systems, along with Poisson noise, there are present other noises. In particular, if dissipation comes from the coupling of the system to a thermal bath, there is white Gaussian thermal noise with intensity $4\Gamma k_BT$, where $T$ is the bath temperature. This noise is described by an extra force $f_T(t)$ in the equation of motion (\ref{eq:Langevin}). Using the explicit form of the characteristic functional for white Gaussian noise, cf. \cite{FeynmanQM}, one can show that such noise leads to an extra term in the function $\psi(k)$ in Eq.~(\ref{eq:distribution_general}), 
\begin{equation}
\label{eq:Gaussian_noise}
(\nu/\Gamma)\psi(k)\to (\nu/\Gamma)\psi (k) + 2 k_BT\omega_0^{-2} k^2.
\end{equation}
In the absence of Poisson noise, $\nu=0$, the distribution $\rho(q)$ (\ref{eq:distribution_general}) is then Gaussian, with no singularities. The distribution is Gaussian also if the noise is Gaussian, but not $\delta$-correlated.

The term $\propto k^2$ in Eq.~(\ref{eq:Gaussian_noise}) imposes an effective cutoff on the values of $k$ that contribute to the integral over $k$ in Eq.~(\ref{eq:distribution_general}), $|k|\lesssim  \omega_0/(k_BT)^{1/2}$. Respectively, the peaks of the distribution $\rho(q)$ are described by the power laws found in Secs.~\ref{sec:singularities} and \ref{sec:underdamped} at distances that exceed $(k_BT)^{1/2}/\omega_0$ from the centers of the peaks. This provides the bound for the observation of the considered power law singularities.

\section{Conclusions}
\label{sec:conclusions}

We have studied the central part of the stationary probability distribution of shot-noise driven systems. We analyzed the probability distributions of the system coordinate and energy. Our central result is that the distributions can display singularities, and in particular peaks, at the equilibrium position of the system as well as away from it. The singularities emerge whether the system dynamics is overdamped or underdamped. They are described by the power law, with characteristic exponents. The positions of the singularities and the exponents are obtained in the explicit form.

The  pattern of the singularities of the stationary distribution provides a reliable indication of the presence of shot noise. In tunable systems, to observe the singularities it is advantageous to make the relaxation rate of the system $t_r^{-1}$ larger than the pulse rate $\nu$. Observing the peak of the coordinate distribution at the equilibrium position requires $\nu t_r<1$, whereas the peaks away from the equilibrium position emerge for $\nu t_r< 1/2$. However, even where $\nu t_r>1/2$, there are still singularities in the derivatives of the distribution. The power-law peak in the energy distribution of underdamped systems at $E=0$ emerges for $\nu t_r < 2$, whereas the peaks for finite energies emerge for $\nu t_r < 2/3$.

The peaks in the probability distributions away from the equilibrium position are characterized by the same exponents, which are determined solely by the parameter $\nu t_r$. The exponents are different for the coordinate and energy distributions. The positions of the peaks, on the other hand, depend on the area of the noise pulses, but are independent of the pulse rate $\nu$. 

Particularly attractive in terms of identifying the noise seem to be underdamped systems, since the distributions can display multiple singularities whose locations form a geometric progression with common ratio $\propto \exp(-\pi\Gamma/\omega_0)$. However, if the system is very strongly underdamped, the singularities may start overlapping, which would complicate resolving them. Also, the characteristic heights of the distribution peaks decrease with the decreasing $\Gamma/\omega_0$, which suggests that $\Gamma/\omega_0$ should not be too small.  This can be of interest for revealing shot noise in side-band cooled vibrational systems, since the cooling leads to the increase of the decay rate \cite{Borkje2010,Purdy2012,Dykman1978,Kippenberg2008}.

The singularities of the probability distribution should occur also in more general types of shot-noise driven systems than the ones discussed above. An important example of current interest is nonlinear vibrational systems modulated by a strong resonant periodic field, with shot noise coming from fluctuations of this field or, in the case of nanomechanical resonators, charge on the resonator. The dynamics of such systems in the rotating frame near the stable state of forced vibrations is described by linearized equations of motion that have a form somewhat different from the standard Langevin equation (\ref{eq:Langevin}) \cite{Zou2012}. However, we expect that the presented results can be extended to such systems and to other systems that perform small-amplitude fluctuations about their stable states.


%

\end{document}